# Pressure induced spin crossover in disordered α-LiFeO$_2$

*Samar Layek [a,\*], Eran Greenberg [a,†], Weiming Xu [a], Gregory Kh. Rozenberg [a], Moshe P. Pasternak,[a] Jean-Paul Itié [b] and Dániel G. Merkel [c,‡]*

[a]*School of Physics and Astronomy, Tel-Aviv University, 69978, Tel-Aviv, Israel*
[b]*Synchrotron SOLEIL, L'Orme des Merisiers, Saint-Aubin, BP 48, 91192 Gif-sur-Yvette Cedex, France*
[c]*European Synchrotron Radiation Facility, F-38043 Grenoble Cedex, France*

[\*]Corresponding author: samarlayek@gmail.com
[†]*Present address:* Center for Advanced Radiation Sources, University of Chicago, Argonne, Illinois 60439, USA
[‡]On leave from Institute for Particle and Nuclear Physics, Wigner Research Centre for Physics, Hungarian Academy of Sciences, H-1525 Budapest, Hungary

Structural, magnetic and electrical-transport properties of α-LiFeO$_2$, crystallizing in the rock salt structure with random distribution of Li and Fe ions, have been studied by synchrotron X-ray diffraction, $^{57}$Fe Mössbauer spectroscopy and electrical resistance measurements at pressures up to 100 GPa using diamond anvil cells. It was found that the crystal structure is stable at least to 82 GPa, though a significant change in compressibility has been observed above 50 GPa. The changes in the structural properties are found to be *on a par* with a sluggish Fe$^{3+}$ high- to low-spin (HS-LS) transition ($S=5/2 \rightarrow S=1/2$) starting at 50 GPa and not completed even at ~100 GPa. The HS-LS transition is accompanied by an appreciable resistance decrease remaining a semiconductor up to 115 GPa and is not expected to be metallic even at about 200 GPa. The observed feature of the pressure-induced HS-LS transition is not an ordinary behavior of ferric oxides at high pressures. The effect of Fe$^{3+}$ nearest and next nearest neighbors on the features of the spin crossover is discussed.

## I. INTRODUCTION

Ferric mono-oxides of the series *A*FeO$_2$ (*A* = Li, Na, and Ag) are well known for their interesting physical properties and practical applications in the field of ion batteries [1,2], gas sensing [3], multiferroics [4,5], gas absorption [6], surface enhanced Raman scattering (SERS) [7], photo-catalyst [8], etc. All these materials can be prepared usually in more than one structure, depending on the preparation conditions. The phase which crystallizes in the rock-salt structure is generally called *alpha phase* (α-*A*FeO$_2$), and an alternative phase with the delafossite-type structure is known as beta phase (β-*A*FeO$_2$) [9]. The α-form of LiFeO$_2$ has the disordered cubic structure (*Fd3m* space group, *a* = 4.158 A [10]) and has been reported in the 1930s [11] as one of the earliest examples of a compound with two different cations randomly distributed on the same crystallographic site. The randomness in the cationic ordering has also been confirmed by both neutron diffraction and magnetic measurements [12,13]. Mössbauer spectroscopy and DC and AC susceptibility measurements show unusual magnetic behavior due to a random distribution of cations in α-LiFeO$_2$ [13]. The ambient conditions Mössbauer spectrum shows a well resolved doublet which can be fit with a distribution of quadrupole electric field gradients reflecting the random distribution in the Fe atoms [13,14]. Tabuchi et al. [15] reported antiferromagnetic order in α-LiFeO$_2$ ($T_N$ ~ 90 K) with an effective moment lower than the theoretical moment typical of high-spin (HS) Fe$^{3+}$ state associated with the formation of short-range antiferromagnetic clusters [12,13]. Also, the frequency dependence of the *ac* peaks at 88 K, the irreversibility behavior, and its evolution with the increasing strength of a superimposed *dc* field, confirm the incompleteness of long range magnetic order suggesting the formation of cluster spin glass [13]. Till today no high-pressure results have been reported for this kind of ferric oxides driving us to investigate the evolution of electronic/magnetic and structural properties in this system.

Applying pressure to such a strongly correlated system usually results in a number of modifications including the quenching of orbital moments, spin crossover, inter-valence charge transfer, insulator-metal transition, moment collapse, and volume collapse [16]. These changes may occur simultaneously or sequentially over a range of pressures. This has been the case of an inter-valence charge transfer observed in the β-polymorph of $CuFeO_2$ [17]. It is noteworthy that many of these changes are the result of the breakdown of the *d* electron localization leading to an insulator-to-metal transition usually concurrent with a collapse of magnetic interactions; namely, the Mott-Hubbard transition [18]. Another cause for the collapse of magnetic moments at high pressure could be spin crossover; a high- to low-spin transition (HS-LS), resulting from the pressure-induced increase of the crystal field [19]. For ferric compounds this will result in a substantial decrease of the magnetic moment ($S=5/2 \rightarrow S=1/2$) and Néel-temperature [20–23], and in complete collapse of magnetism in ferrous compounds ($S = 2 \rightarrow S = 0$). The latter case has been observed in several ferrous oxides such as wüstite (FeO) [24], $Mg_{1-x}Fe_xO$ [25] and FeS [26]. For some systems the closure of the Mott-Hubbard gap could be driven by a HS-LS transition [27,28]. For ferric compounds, the HS-LS transition is usually a first-order transition observed at the 40 – 60 GPa range. It is noteworthy, that for all studied ferric compounds mentioned above, all iron magnetic moments had an identical neighborhood. Contrary, α-LiFeO$_2$, similar to some ferrous systems (e.g. $Mg_{1-x}Fe_xO$ [25,29,30]), is characterized by the random environments of the Fe ions. It is expected that such a feature may have significant impact on the evolution of the electronic and structural properties of α-LiFeO$_2$ under pressure.

In the present study we attempted to reveal and analyze these features applying thorough high pressure studies of structural, magnetic and electrical transport properties of α-LiFeO$_2$ by means of Synchrotron X-ray diffraction (SXRD), $^{57}$Fe Mössbauer spectroscopy (MS) and electrical resistance measurements.

**II. EXPERIMENTAL TECHNIQUES**

A polycrystalline sample of α-LiFeO$_2$ was prepared by the solid state reaction method as reported in [31]. Stoichiometric amounts of high purity $Li_2CO_3$ and $Fe_2O_3$ were mixed, pelletized and heated at 800 °C for 2 hours in air atmosphere. Another batch of sample containing 25% enriched $^{57}$Fe was also prepared for high-pressure MS experiments.

High-pressure SXRD measurements were performed at the Psiché beamline, Synchrotron SOLEIL, France, at room temperature (RT) in angle-dispersive mode (λ = 0.3738 Å) with patterns collected using a MAR detector and integrated using the FIT2D [32] and DIOPTAS [33] programs. A few of the SXRD data were collected at ESRF, Grenoble, at the ID-27 beam line using the same wavelength. The results were analyzed by Rietveld refinement using the GSAS [34] and EXPGUI packages [34,35]. Diamond anvil cells (DACs) manufactured at Tel-Aviv University [36] with diamond anvils with culet diameters of 400 and 200 μm were used up to 15 and 82 GPa, respectively. Re gaskets with a starting thickness of 250 μm were pre-indented to 30 and 15 μm, and holes of 180 and 100 μm diameters were drilled at the center of the indentation, for cells of the larger and smaller culet sizes, respectively. Samples, along with spherical ruby chips, were placed at the center of the drilled cavity. Nitrogen and helium were used as pressure transmitting medium for the 400 and 200 μm anvil cells, respectively.

High-pressure MS measurements were carried out with a DAC with diamond anvil culets of 250 μm, prepared in the same manner mentioned above. $N_2$ was used as a pressure medium. A $^{57}$Co(Rh) point-source with an initial activity of 10 mCi was used in the transmission geometry. Low-temperature measurements down to 8 K were performed using a custom made top-loading liquid nitrogen-helium cryostat. MS spectra were fit using the least-squares fitting method (MossA) to obtain the MS hyperfine parameters, namely, the relative abundance of the components, the isomer shift (IS), electric quadrupole splitting (QS) and magnetic hyperfine field $H_{hf}$ [37]. Low temperature measurements at 76 and 100 GPa down to 2.2 K were recorded using synchrotron Mössbauer spectroscopy (SMS) measurements at the ID18 beamline of ESRF using the techniques described in [38] and helium pressure medium. Velocity values obtained from the SMS measurements are affected by the second-order Doppler effect since the source is kept at RT, while the sample is cooled down.

High pressure electrical resistance measurements up to 120 GPa were performed with 200 μm culet anvils. The Re gasket was covered with an insulating layer of an $Al_2O_3$-NaCl mixture (3:1 atomic ratio), which also serves as the pressure medium. Samples with ruby chips were placed inside a 100 μm cavity drilled within the pressed insulating layer. Platinum foils with a thickness of 5-7 μm were cut in triangular form and used as electrical probes for resistance measurements. The Pt foils were connected to copper leads, at the base of the diamond anvil, using a silver epoxy. At each pressure, under both compression and decompression cycles, resistance was measured as a function of temperature using a standard four-probe method in a custom-made cryostat. At each temperature, the voltage was measured as a function of a series of applied currents, for determining the resistance from the obtained slope.

Pressure was measured both before and after each measurement from the ruby fluorescence spectra [39,40,41]. Diamond Raman spectra were also used to determine the pressure for XRD and resistance measurements [42], especially at pressures above 60 GPa.

## III. RESULTS

### X-ray diffraction

Synchrotron powder XRD measurements were performed up to 82 GPa. Up to pressures around 12 GPa they were carried out using $N_2$ as pressure transmitting medium, whereas above 12 GPa to around 82 GPa using He pressure medium. XRD patterns in the compression cycle are shown in Figure 1. All the peaks could be identified to arise from the original fcc structure (space group $Fm3m$). The structure is stable up to the highest pressure. As can be seen, the XRD peaks shift to higher angle due to the decrease in the unit-cell volume.

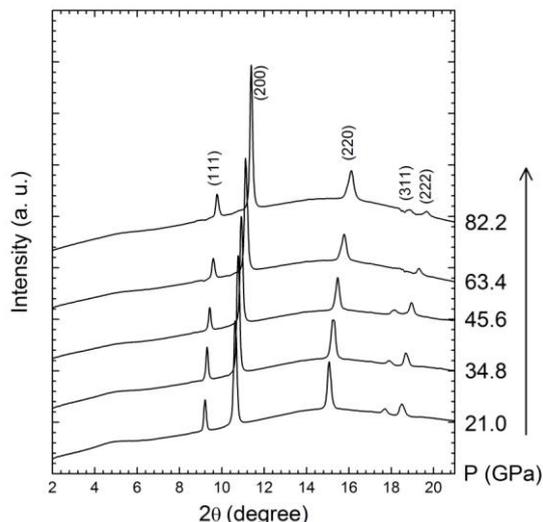

Figure 1. Synchrotron XRD pattern of $LiFeO_2$ at various pressures. The crystal structure remains the same up to the highest applied pressure (λ=0.3738 Å).

A typical Rietveld refinement of the XRD pattern at 21.0 GPa is shown in Figure 2. For the fit, $Li^{1+}$ and $Fe^{3+}$ cations at (0, 0, 0) octahedral sites and oxygen anions at (½,½,½) were set as the initial atomic positions. The extracted crystal volume values are plotted in Figure 3 as a function of pressure. An abrupt change in the slope of $V(P)$ is observed at around 50 GPa (see Figure 3). The data for the molar volume below and above 50 GPa can be fit with two different second-order Brich-Murnaghan (BM2) equations of state EOS [43]. The performed fit results in $K_0$=152.8(7), $V_0$=71.31(4) below 50 GPa, and $K_{50}$ = 334.7(2), $V_{50}$ = 57.6(1) and $K_{50}$ = 192.0(4) GPa, $V_{50}$ = 57.1(7) Å$^3$ for the pressure regions below and above 50 GPa, respectively. There $K_0$, $V_0$ and $K_{50}$, $V_{50}$ are the bulk moduli , and the unit-cell volumes at 1 bar and 50 GPa (300 K), correspondingly. The drastic change of the compressibility at about 50 GPa suggests a sluggish electronic transformation above this pressure.

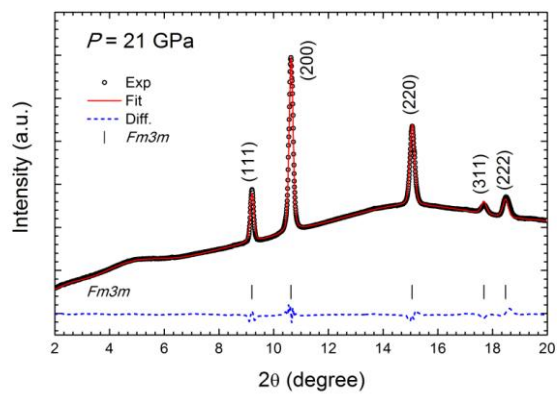

Figure 2. (Color online) Rietveld refinement of the SXRD pattern of $LiFeO_2$ at 21.0 GPa (λ=0.3738 Å). Black circles are the experimental points, solid red line is the fit, black bars are Bragg positions assuming the $Fm3m$ space group, and difference between experimental data and theoretical fit is shown using a dashed blue line.

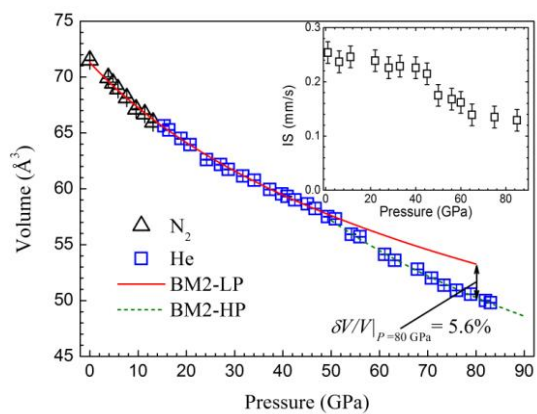

Figure 3: (Color online) Unit-cell volume of α-$LiFeO_2$ as a function of pressure. Black triangles and blue squares are the data obtained with nitrogen and helium pressure medium, respectively. Solid red and dashed green lines represent the 2nd order Birch-Murnaghan fit for the data below and above 50 GPa, respectively. The inset shows the pressure dependence of the isomer shifts extracted from the RT Mössbauer spectra (IS values are relative to α-Fe at RT).

*Mössbauer Spectroscopy*

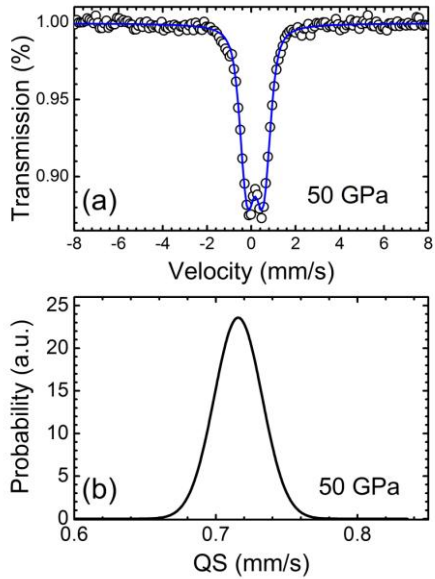

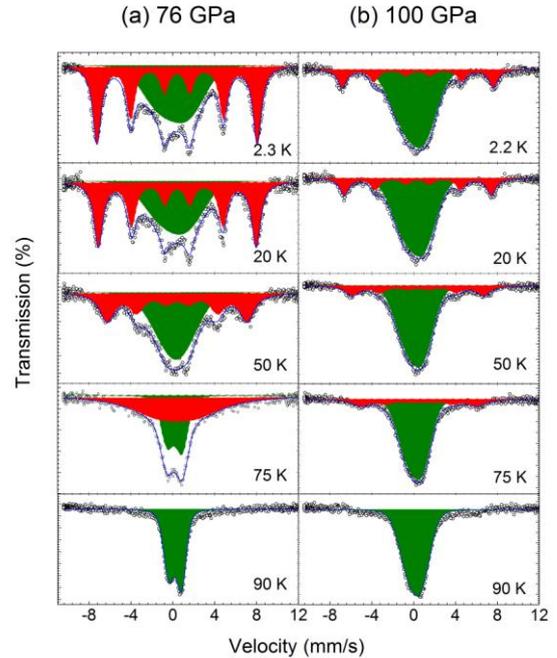

**Figure 4:** (Color online) (a) Typical RT Mössbauer spectrum (black circles) recorded at 50 GPa and room-temperature along with the fit (solid line) assuming the quadrupole splitting distribution function shown in (b). Velocity values are with respect to α-Fe at RT.

**Figure 5:** (Color online) Mössbauer spectra recorded using synchrotron Mössbauer spectroscopy at (a) 76 and (b) 100 GPa at various temperatures. Velocity values are relative to α-Fe at RT. Black circles are the experimental data. Green, red and blue solid lines are the non-magnetic quadrupole doublet, high spin magnetic sextet and total fit of the data, respectively. The antiferromagnetic ordering temperature remains nearly 90 K for both the pressure.

LiFeO$_2$ is antiferromagnetic below 90 K. At RT, the Mössbauer spectrum consists of what appears to be a paramagnetic (non-magnetic) doublet. The spectrum at ambient conditions is best fit with a quadrupole splitting distribution instead of a single doublet. This can be explained by the disorder associated with the fact that Li and Fe cations randomly occupy the same crystallographic position [13]. Based on this assumption a probability distribution of quadrupole splitting can be extracted from the experimental spectra. The RT spectrum remains nearly the same up to the highest applied pressure around 100 GPa. A typical RT Mössbauer spectrum fit and corresponding quadrupole distribution is plotted in Figure 4 (a) and (b), respectively. The inset of Figure 3 shows the variation of the isomer shift (IS), derived from the Mössbauer spectra fits, as a function of pressure. The IS decreases slowly with pressure up to around 50 GPa, at which point a significant decrease is found.

Low-temperature MS spectra were collected in order to determine the electronic state of Fe$^{3+}$ at high pressures and the origin of the unusual drop in IS(*P*). At *T*<,*T$_N$* (90 K) Mössbauer spectra at ambient pressure can be fit with a magnetic sextet corresponding to the HS Fe$^{3+}$ magnetic state. This behavior does not change up to about 50 GPa, above which a non-magnetic component persists even at *T*<90 K. Figures 5 (a) and (b) show two such examples of low temperature measurements at 76 and 100 GPa, respectively. Above 50 GPa, the spectrum below 90 K can only be fit with a combination of a magnetic sextet and a quadrupole distribution component. The calculated quadrupole distribution functions at *P*=100 GPa are shown in Fig. 6(a). The calculated relative abundance *vs.* temperature of the magnetically split component at 76 and 100 GPa is plotted in Fig. 6(b). The fact that at low temperatures this abundance reaches a saturation value means that the significant amount of the component attributed to the quadrupole distribution at low temperatures is not due to a lower ordering temperature but rather due to an onset above 50 GPa of a new nonmagnetic high pressure (HP) component. This new component is characterized by a significantly reduced IS value; the difference between IS values of the low-pressure and high-pressure phase components is ~0.24 mm/s at 76 GPa and 0.29 mm/s at 100 GPa, respectively. The HP component also shows two special features of the distribution function; at the same pressure, lowering the temperature results both in a higher mean quadrupole splitting value and in a broader distribution function. It is noteworthy, that from the RT spectra one cannot distinguish between low- and high-pressure components due to the large quadrupole distribution; therefore, the IS value obtained at RT above 50 GPa is in fact an average of both components.

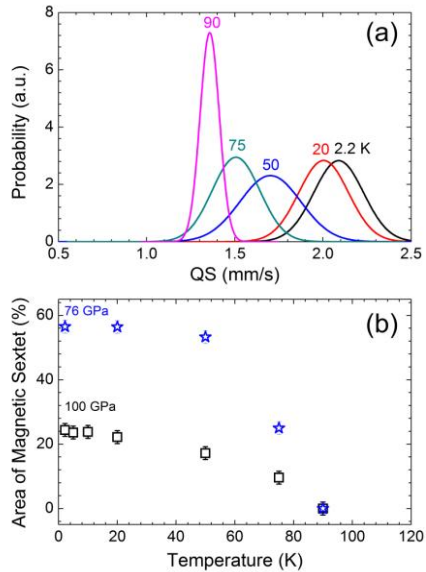

**Figure 6:** (Color online) (a) Quadrupole distribution fit for Mössbauer spectra at 100 GPa for different temperatures and (b) area of the magnetic sextet component as a function of temperature at 76 and 100 GPa.

*Electrical Resistance Measurements*

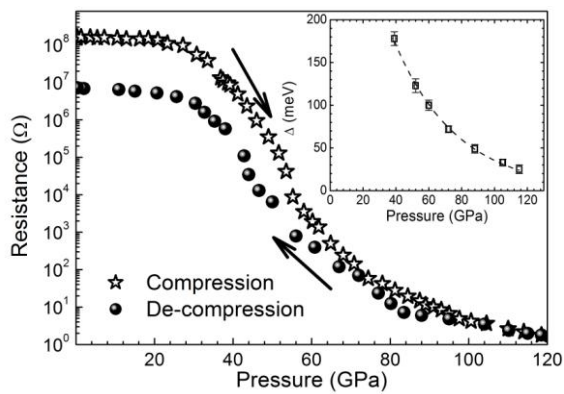

**Figure 7:** Room-temperature resistance of α-LiFeO$_2$ as function of pressure during compression (stars) and decompression (spheres) cycles. The inset shows the variation of the activation energy as a function of pressure. Dashed line is the fit according to $\Delta(P) = \Delta_0 \exp(-\alpha P)$ equation showing an exponential decrease of the activation energy. The large change in the resistance between the compression and decompression cycles is explained as due to compacting of the powder sample.

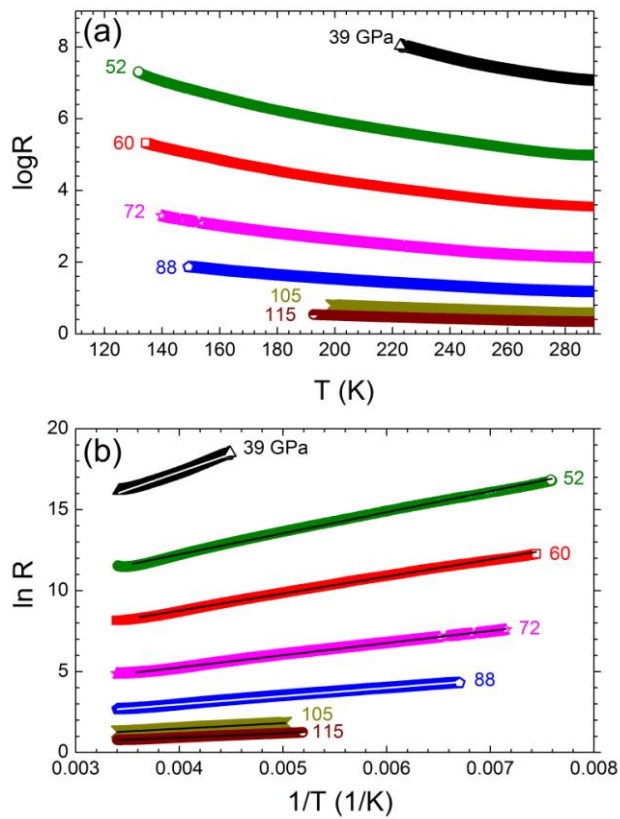

**Figure 8:** (Color online) (a) Variation of resistance as a function of temperature at various pressures. (b) ln(R) as a function of inverse temperature where solid lines are the linear fit according to the Arrhenius equation.

Room-temperature resistance as a function of pressure for both compression and decompression cycles are shown in Figure 7. During compression, up to 30 GPa the RT resistance remains nearly constant. Above 30 GPa, a continuous decrease in the resistance can be seen up to the highest pressure studied (120 GPa). The steepest decrease in the resistance occurs at ~50 GPa. Fig. 8a shows the variation of resistance as a function of temperature at different pressures from 39 to 115 GPa. Below 39 GPa, R(T) was not possible to obtain due to the high value of the resistance. Throughout the whole pressure range, the resistance increases with decreasing temperature indicating a semiconducting behavior. The temperature dependence of the resistance can be fit according to the Arrhenius equation $R = R_0 \exp(\Delta/2k_B T)$ where $\Delta$ is the activation energy for the electrical transport and $k_B$ is the Boltzmann constant (Fig. 8b). The inset of Figure 7 shows the activation energy, calculated by the above expression, as a function of pressure. Similar to the RT resistance, we observe a continuous decrease in the activation energy with increasing pressure. However, no signature of an insulator-metal transition has been found at least up to the highest pressure studied ~120 GPa.

Pressure dependence of the energy gap can be fit with the expression $\Delta(P) = \Delta_0 \exp(-\alpha*P)$ where $\Delta_0 = 486$ meV and $\alpha = 0.026$ GPa$^{-1}$ are the ambient pressure energy gap and rate at which the gap closes, respectively. The observed nonlinearity of the pressure dependence of the energy gap may result from the nonlinearity of the compressibility of the material (see [44]). The value of the band gap calculated at 200 GPa is ~ 2.7 meV. Thus, no insulator to metal transition can be expected at least up to 200 GPa, assuming no prior structural transition.

## IV. DISCUSSIONS

Summarizing the MS and XRD results, we can conclude that in α-LiFeO$_2$, similar to other ferric compounds [16,45], an onset of electronic transition is observed at about 50 GPa. This electronic transition is characterized by an appearance of a new, non-magnetic MS component, characterized by a significantly reduced isomer shift value. Low-temperature measurements show the gradual broadening of the absorption spectrum and the increase of QS with the decrease in temperature. All these features are indicators of a transition to a LS state. The significant broadening of the absorption spectra at low temperatures is a feature of a paramagnetic spin relaxation phenomenon, a typical characteristic of the Fe$^{3+}$ LS state (see [20]). Therefore, we can determine that the electronic transition is a HS-LS, and that with increasing pressure, the abundance of the LS component increases. The increase of the abundance of the LS state corroborates with the sluggish crystal volume decrease attributed to a significant (~12%) decrease in the FeO$_6$ polyhedral volume resulting from the transition to the LS state [46]. This transition is obviously second-order starting at 50 GPa, yet not completed even at ~100 GPa. In contrast to LiFeO$_2$, in the previously studied ferric oxides [16,45] and references therein] a first-order HS-LS transition associated with a precipitous volume reduction was observed at the 40 – 60 GPa range. Such significant difference in the features of the Fe$^{3+}$ spin crossover can be associated with the different environment of Fe$^{3+}$ ions in the above-mentioned systems. Indeed, in α-LiFeO$_2$ a random distribution of iron cations should result in significant inhomogeneity of the crystal-field splitting distribution and therefore different critical pressure values for the transition to the LS state. Thus, our claim is consistent with recent studies of some ferrous compounds, namely: Mg$_{1-x}$Fe$_x$O [25,29,30], (Mg$_{0.9}$Fe$_{0.1}$)$_2$SiO$_4$ olivine [47] and ringwoodite (Mg$_{1-x}$Fe$_x$)$_2$SiO$_4$ [48]. In all these compounds, characterized by a random Fe$^{2+}$ environment, a sluggish spin crossover phenomenon was observed with the range of transition of about 20-30 GPa. In the case of α-LiFeO$_2$ the spin crossover range is even more (at least 60 GPa), which could be related with more significant difference in ionic radii sizes of the cations in the present case.

It is noteworthy, that recent theoretical calculations [27, 49] show that the HS-LS crossover in $d^5$ systems strongly suppresses the effective Hubbard parameter $U_{eff}$, resulting in an abrupt significant decrease of resistance [e.g. 20] and in some cases in the complete closure of the Hubbard gap [23, 28]. In the present case, we observed an appreciable resistance reduction of about eight orders of magnitude, which takes place within a rather broad pressure range coinciding with the sluggish HS-LS transition. However, despite of such appreciable resistance decrease, the material remains semiconductor up to 115 GPa, the highest pressure measured, and is not expected to be metallic even at about 200 GPa.

## V. CONCLUSIONS

In conclusion, despite that the crystal structure α-LiFeO$_2$ is stable at least to 82 GPa, a significant increase in compressibility was observed above 50 GPa. This change is the result of the sluggish second-order Fe$^{3+}$ high- to low-spin transition which starts at 50 GPa and not completed even at ~100 GPa. The HS-LS transition coincides with the appreciable resistance decrease, however, no gap closure was observed up to 115 GPa. The observed feature of the pressure-induced HS-LS transition in α-LiFeO$_2$ contradicts with the ordinary behavior of ferric oxides and is caused, presumably , by the random environment of Fe$^{3+}$ ions. Additional studies of Li ferrites, particularly in the ordered polymorphs of LiFeO$_2$ and LiFe$_5$O$_8$, (which can be also prepared in both ordered and disordered phases), is desirable to clarify the mechanism of the studied electronic transition and the effect of the nearest Fe$^{3+}$ environment upon the transition features.

**Acknowledgements**

Mark Shulman and Davide Levy are sincerely acknowledged for his help in the initial stages of few experiments. A few of the SXRD data were collected at the ID-27 beamline of the European Synchrotron Radiation Facility, Grenoble, France. We are grateful to Volodymyr Svtlyk and Andrew Cairns at the ESRF for providing assistance in using beamline ID-27. This research was supported by the Israeli Science Foundation Grant #1489/14.